\documentclass[twocolumn,tighten]{aastex61}
\usepackage{amssymb, amsmath,framed}

\usepackage{threeparttable}
\usepackage{color}

\usepackage{bm}
\expandafter\ifx\csname package@font\endcsname\relax\else
 \expandafter\expandafter
 \expandafter\usepackage
 \expandafter\expandafter
 \expandafter{\csname package@font\endcsname}
\fi
\def\bq{\begin{equation}}
\def\eq{\end{equation}}
\def\bqy{\begin{eqnarray}}
\def\eqy{\end{eqnarray}}
\def\calo{\mathcal{O}}

\begin{document}
\title{Is Proxima Centauri b habitable? -- A study of atmospheric loss}

\correspondingauthor{Chuanfei Dong}
\email{dcfy@princeton.edu}

\author{Chuanfei Dong}
\affiliation{Department of Astrophysical Sciences and Princeton Plasma Physics Laboratory, Princeton University, Princeton, NJ 08544, USA}

\author{Manasvi Lingam}
\affiliation{Harvard John A. Paulson School of Engineering and Applied Sciences, Harvard University, Cambridge, MA 02138, USA}
\affiliation{Harvard-Smithsonian Center for Astrophysics, The Institute for Theory and Computation, 60 Garden Street, Cambridge, MA 02138, USA}

\author{Yingjuan Ma}
\affiliation{Institute of Geophysics and Planetary Physics, University of California, Los Angeles, CA 90095, USA}

\author{Ofer Cohen}
\affiliation{Lowell Center for Space Science and Technology, University of Massachusetts, Lowell, MA 01854, USA}

\begin{abstract}
We address the important question of whether the newly discovered exoplanet, Proxima Centauri b (PCb), is capable of retaining an atmosphere over long periods of time. This is done by adapting a sophisticated multi-species MHD model originally developed for Venus and Mars, and computing the ion escape losses from PCb. The results suggest that the ion escape rates are about two orders of magnitude higher than the terrestrial planets of our Solar system if PCb is unmagnetized. In contrast, if the planet does have an intrinsic dipole magnetic field, the rates are lowered for certain values of the stellar wind dynamic pressure, but they are still higher than the observed values for our Solar system's terrestrial planets. These results must be interpreted with due caution, since most of the relevant parameters for PCb remain partly or wholly unknown.
\end{abstract}

\section{Introduction} \label{SecIntro}
Currently, thousands of exoplanets with diverse characteristics have been detected \citep{WF15}. Arguably, one of the most important reasons for studying them is to determine whether they could be habitable. The Habitable Zone (HZ) is broadly defined as the region around a star where a planet can support liquid water for certain values of the atmospheric pressure \citep{Letal09}, and has been extensively investigated \citep[e.g.][]{KWR93,KC03,Kop13}. As per current estimates, it is expected that there are $\sim 10^{10}$ Earth-sized planets that lie within the HZ of their stars, in the Milky Way \citep{MWP,WF15}. 

Most of the attention has focused on HZ studies of exoplanets around M-dwarfs for two reasons: (i) Earth-sized planets in the HZ are quite common \citep{DC15}, and (ii) detailed characterizations of their atmospheres are easier since they are situated much closer to the host star \citep{SD10,SBJ16}. Hence, there have been numerous HZ studies of M-dwarf exoplanets \citep{Tart07,Scal07,SBJ16}. This area received a major impetus recently with the potential discovery of a super-Earth (a planet slightly larger than the Earth) orbiting Proxima Centauri \citep{AEA16}, which was christened Proxima Centauri b (PCb). The result is significant because Proxima Centauri is the nearest star to the Earth, and plans are already underway for exploring PCb.\footnote{\url{https://breakthroughinitiatives.org/Initiative/3}}

Exoplanetary atmospheres are an important area of research since spectroscopic analyses can potentially reveal the presence of life via biosignatures \citep{SD10}. This raises an important question: at what rates are the atmospheres of close-in exoplanets (such as those in the HZ of M-dwarfs) eroded by the stellar wind? The question is of paramount importance, as high escape rates might deplete the planet of its atmosphere after a short period of time, if one excludes mechanisms such as outgassing \citep{KC03}.

In our own Solar system, the importance of atmospheric loss mechanisms has been thoroughly documented for Mars \citep[e.g.][]{Lammer13,Lil15,dong15b}. One of the primary scientific goals of NASA's Mars Atmosphere and Volatile EvolutioN (MAVEN) mission is to determine the escape rates, which are crucial because of their potential impact on the long-term evolution of the Martian atmosphere (e.g., loss of water). \citet{brain15} estimated a net ion escape rate of $\sim$2.5 $\times$ 10$^{24}$ s$^{-1}$ by choosing a spherical shell at $\sim$1000 km above the planet with energies $>$25 eV over a 4-month MAVEN period. The ion loss increases by more than one order of magnitude during interplanetary corona mass ejection events \citep{dong15a,jakosky15b}.

There exist several papers that have explored atmospheric losses from HZ exoplanets \citep[e.g.][]{Khod07,Tian09,Erk13}, some of which relied upon \emph{hydrodynamic} models. As we have learned from our Solar system, realistic estimates appear to necessitate sophisticated \emph{magnetohydrodynamical} (MHD) models for studying the stellar wind-exoplanet interaction, especially for Earth-sized (or above) planets where neutral losses (e.g., thermal and photochemical escape) are not the dominant mechanism \citep{brain16}. Such studies have been quite sparse in exoplanetary research as they necessitate highly detailed numerical codes. However, recent publications \citep[e.g][]{Coh14,KJ14,CMD15} on magnetized and unmagnetized exoplanets are notable for tackling this issue. 

In this Letter, we investigate atmospheric losses arising from stellar wind interaction with a planet that has PCb-like parameters. Since PCb may lack a magnetic field \citep{Barn16}, we examine both the magnetized and unmagnetized cases. Although our final results are outwardly simple, we note that this interaction process is actually highly complex in nature. Our model self-consistently accounts for the ionospheric chemistries and electromagnetic forces \citep{Ma04,Ma13}.

\section{The simulation setup}
In this Section, we describe our code, and the rationale behind our choices of the parameters. 

\subsection{Physical model and computational methodology}
The 3-D Block Adaptive Tree Solar-wind Roe Upwind Scheme (BATS-R-US) multi-species MHD (MS-MHD) model was initially developed for Mars \citep{Ma04} and Venus \citep{Ma13}. The latter is employed herein, and the neutral atmospheric profiles are based on the solar maximum conditions. The MS-MHD model solves a separate continuity equation for each ion species, and one momentum and one energy equation for the four ion fluids H$^+$, O$^+$, O$_2^+$, CO$_2^+$ \citep{Ma04,Ma13}. In contrast to most global magnetosphere models applied to Earth that start from $2$-$3$ Earth radii, the Mars/Venus MS-MHD model contains a self-consistent ionosphere, and thus the lower boundary extends down to 100 km altitude above the planetary surface. The MS-MHD model includes ionospheric chemical processes such as charge exchange and photoionization.

The chemical reactions and rate coefficients are based on \citet{schunk2009}, except the photoionization rates; the latter are rescaled to PCb values based on the EUV estimate in \citet{Rib16}. The O$^+$, O$_2^+$, CO$_2^+$ ion densities at the model lower boundary (i.e., on the 100 km altitude sphere) satisfy the photochemical equilibrium condition, where float boundary conditions for the velocity $\mathbf{u}$ and the magnetic field $\mathbf{B}$ have been applied. Given the high collision frequency near the inner boundary, ions, electrons and neutrals are expected to have roughly the same temperature. Thus, the plasma temperature (sum of ion and electron temperatures) is twice the neutral temperature. 

As this code is capable of handling a wide array of chemical and physical processes, and has been thoroughly benchmarked and validated for Venus and Mars, we use it for studying the ion escape rates of PCb. As a consequence, we assume that the atmospheric composition of PCb is similar to that of Venus and Mars. In reality, we wish to note there are many possibilities for the atmospheric composition of PCb, provided that it even exists \citep{Rib16,Turb16,Gold16}. A clearer picture regarding the existence and composition of the atmosphere may emerge via the JWST \citep{KL16}.

In the model, a nonuniform, spherical grid is adopted in order to accurately capture the multi-scale physics in different regions. The radial resolution varies from 5 km ($\sim 1/2$ scale height) at the inner boundary to several thousands of km at the outer boundary. The horizontal resolution is $3.0^{\circ}$ (in both longitude and latitude). The simulation domain is defined by $-$45 R$_P \leq$ X $\leq$ 15 R$_P$ and $-$30 R$_P \leq$ Y, Z $\leq$ 30 R$_P$, where R$_P$ denotes the planetary radius. The code is run in the PCb-Star-Orbital (PSO) coordinate system, where the $x$ axis points from PCb toward Proxima Centauri, the $z$ axis is perpendicular to the PCb's orbital plane, and the y axis completes the right-hand system. 

Table \ref{table1} summarizes the chemical reaction schemes and the associated rates for inelastic collisions used in the MS-MHD calculations.

\begin{table}
\caption{Chemical reactions and associated rates (the reaction rates are adopted from \citealt{schunk2009}).}\label{table1}
\centering
\begin{tabular}{lll}
\hline
\hline
\multicolumn{2}{c} {Chemical Reaction} & Rate Coefficient \footnote{Electron impact ionization is neglected in the calculation, H$^+$ density is from the stellar wind, the neutral hydrogen is neglected.} \\
\hline
\multicolumn{3}{c} {Primary Photolysis and Particle Impact\footnote{The photoionization frequencies are rescaled to PCb values, using the EUV estimate from \citet{Rib16}, the latter being around 33 times that received by the Earth.} in s$^{-1}$} \\
\hline
& CO$_{2}$ + $h\nu$ $\rightarrow$  CO$_2^+$ + $e^{-}$ &  5.55 $\times$ 10$^{-5}$   \\ %(solarmin)$
& O + $h\nu$ $\rightarrow$  O$^+$ + $e^{-}$     &  2.08 $\times$ 10$^{-5}$  \\
\hline
\multicolumn{3}{c} {Ion-Neutral Chemistry in cm$^3$ s$^{-1}$} \\
\hline
&  CO$_{2}^+$ + O $\rightarrow$  O$_2^+$ + CO  & $1.64 \times 10^{-10}$   \\
&  CO$_{2}^+$ + O $\rightarrow$  O$^+$ + CO$_2$  & $9.60 \times 10^{-11}$  \\
&  O$^{+}$ + CO$_2$ $\rightarrow$  O$_2^+$ + CO  & $1.1 \times 10^{-9}$ (800/T$_i$)$^{0.39}$  \\
&  H$^{+}$ + O $\rightarrow$  O$^+$ + H \footnote{Rate coefficient from \citet{fox01}.} & $5.08 \times 10^{-10}$  \\
\hline
\multicolumn{3}{c} {Electron Recombination Chemistry in cm$^3$ s$^{-1}$} \\
\hline
&  O$_2^{+}$ + $e^{-}$ $\rightarrow$  O + O  & $7.38 \times 10^{-8}$ (1200/T$_e$)$^{0.56}$ \\
&  CO$_2^{+}$ + $e^{-}$ $\rightarrow$  CO + O  & $3.10 \times 10^{-7}$ (300/T$_e$)$^{0.5}$ \\
\hline
\hline
\end{tabular}
\end{table}

\begin{table}[htp]
\caption{The stellar wind input parameters at PCb \citep{GDC16} for two case studies in the PSO coordinate system. Case 1 (C1) corresponds to the maximum P$_{dyn}$ and P$_{tot}$ over one orbital period of PCb. Case 2 (C2) corresponds to minimum P$_{dyn}$ and P$_{tot}$, but with the maximum P$_{mag}$.} \label{SW}
\centering
\begin{tabular}{lllll}
\hline
\hline
 & n$_{sw}$ (cm$^{-3}$) & T$_{sw}$ (K) & v$_{sw}$\footnote{The $y$-component of the velocity arises primarily from the orbital motion of PCb.} (km/s) & IMF (nT) \\
\hline
C1 & 21400 & 8.42$\times$10$^5$ & (-833, 150, 0)  & (0, 0, -227)  \\
\hline
C2 & 2460 & 9.53$\times$10$^5$ &  (-1080, 150, 0) & (0, 0, -997) \\
\hline
\hline
\end{tabular}
\end{table}

\subsection{The selection of the simulation parameters}
We adopt stellar and planetary parameters that are consistent with PCb. However, we wish to caution the readers that our exoplanet is \emph{not} necessarily representative of PCb itself, as many of these parameters are partly or wholly unknown.

The Venusian neutral atmospheric profile is used as a prototype since PCb may experience significant losses of water and $H_2$, which are potentially major components, quite rapidly \citep{Rib16,Barn16}. As a result, the overall composition could eventually resemble that of Venus. It may also be relatively more suited to modeling the short distance ($\sim 0.05$ AU) between Proxima Centauri and PCb \citep{AEA16}. But, we note that the Venusian atmosphere is possibly denser (and colder) compared to PCb. Both of these factors will alter the scale height, which in turn affects the escape rates. The Venusian neutral atmosphere from \citet{Ma13} is: 
\begin{eqnarray} \label{Vatm93} 
[CO_2] &=& 1.0 \cdot 10^{15} \cdot e^{-(z-z_0)/5.5} ~cm^{-3}, \\ \nonumber
[O] &=& 2.0 \cdot 10^{11} \cdot e^{-(z-z_0)/17} ~cm^{-3}.
\end{eqnarray}
where $z_0=100$ km. We reconstruct the neutral atmospheric profiles based on the following parameters. First, the study by \citet{AEA16} concluded that its \emph{minimum} mass was $1.27\,M_\oplus$, which yields a minimum radius of around $1.1\,R_\oplus$ \citep{AEA16} assuming a rocky composition akin to that of the Earth, i.e., adopting the empirical relationship $R/R_{\oplus} \sim (M/M_{\oplus})^{1/3.7}$ \citep[e.g.][]{valencia06}. If we assume that only the gravity changes, the scale height of PCb is
\begin{equation} \label{ScHPCB}
    H_{\mathrm{PCb}} = \frac{kT}{mg} = 0.85\, H_{\mathrm{Venus}},
\end{equation}
and we will express all quantities in terms of the Venusian values detailed in \citet{Ma13}, since the code was calibrated for Venus. The second parameter required as an input is the density at the base of the model, which requires the surface atmospheric pressure, p$_{sur}$. A precise value of p$_{sur}$ remains unknown at this stage - we choose the Earth value of $1$ bar. A similar value was also considered in the simulations by \citet{Mead16}. Thus, we obtain a density that is $0.011$ times the Venus value at the model lower boundary, the reduced value primarily arising from the much lower atmospheric pressure. PCb's neutral atmospheric profiles are as follows:
\begin{eqnarray} \label{Vatm1} 
[CO_2] &=& 1.1 \cdot 10^{13} \cdot e^{-(z-z_0)/4.7} ~cm^{-3} \\ \nonumber
[O] &=& 2.2 \cdot 10^{9} \cdot e^{-(z-z_0)/14.5} ~cm^{-3}
\end{eqnarray}
It is noteworthy that the density of the dominant neutral component at 100 km altitude on the Earth (N$_2$) and PCb (CO$_2$) are about 1$\times$10$^{13}$ cm$^{-3}$, nearly equal to one another. Furthermore, the scale height of N$_2$ at 100 km is about 5.8 km which is close to the scale height of CO$_2$ for Venus at 100 km, before rescaling it to account for the higher gravity of PCb. Therefore, the reconstructed neutral atmosphere mimics certain features of Earth's atmosphere. The sensitivity of our results to the atmospheric pressure is evaluated by also considering the case with Venusian surface pressure ($93$ bar).

Next, the stellar wind parameters at PCb must be specified, namely the density $\left(n_{sw}\right)$, temperature $\left(T_{sw}\right)$, velocity $\left(v_{sw}\right)$ and interplanetary magnetic field ($\mathrm{B_{IMF}}$). The dynamic pressure $P_{dyn} = m_p n_{sw} v_{sw}^2$ ($m_p$ is proton mass), magnetic pressure $P_{mag} = \mathrm{B_{IMF}^2}/(2\mu_0)$ and total pressure $P_{tot} = P_{dyn}+P_{mag}$ are also introduced. The relevant values have been listed in Table \ref{SW}, and have been chosen from \citet{GDC16} with either maximum or minimum total pressure (and dynamic pressure), over one orbit of PCb. The stellar wind MHD model is driven at its inner boundary by the observable stellar surface magnetic field (i.e., magnetogram).

For the magnetized case, we have chosen PCb's dipole moment to be approximately one-third of the Earth with the same dipole polarity. A similar value has also been adopted for PCb in recent studies \citep{ZB16}. As PCb rotates slower than the Earth \citep{Rib16}, this choice is consistent with certain dynamo scaling laws \citep{Christ10}.

\section{The Simulation Results}

\begin{table}
\caption{Ion escape rates in sec$^{-1}$.}\label{table3}
\centering
\begin{tabular}{lllll}
\hline
& O$^+$ & O$_2^+$ & CO$_{2}^+$ & Total  \\
\hline
\multicolumn{5}{c} {PCb with 1 bar surface pressure} \\
\hline
C1-UnM\footnote{Unmagnetized Case 1.} & 1.8$\times$10$^{27}$  & 2.4$\times$10$^{26}$   &  3.3$\times$10$^{26}$   &  2.4$\times$10$^{27}$ \\ %85.30 kg/sec
C2-UnM & 1.1$\times$10$^{27}$  &  9.5$\times$10$^{25}$  &   8.2$\times$10$^{25}$  &  1.3$\times$10$^{27}$ \\ %40.56 kg/sec
C1-M\footnote{Magnetized Case 1.} & 7.3$\times$10$^{26}$  &  5.4$\times$10$^{26}$  &   5.8$\times$10$^{26}$  &  1.8$\times$10$^{27}$ \\ %91.12 kg/sec
C2-M & 5.9$\times$10$^{25}$  &  8.7$\times$10$^{25}$  &   5.3$\times$10$^{25}$  &  2.0$\times$10$^{26}$ \\ %10.14 kg/sec
\hline
\multicolumn{5}{c} {PCb with 93 bar surface pressure\footnote{Scale height, H$_{\mathrm{PCb}}$, is still equal to 0.85 H$_{\mathrm{Venus}}$.}} \\ %99.25 kg/sec
\hline
C1-UnM$_{93}$ & 3.7$\times$10$^{27}$  & 4.1$\times$10$^{24}$   &  1.4$\times$10$^{23}$   &  3.7$\times$10$^{27}$ \\ 
\hline
\end{tabular}
\end{table}

\begin{figure*}[!ht]
\centering
\includegraphics[scale=1.0]{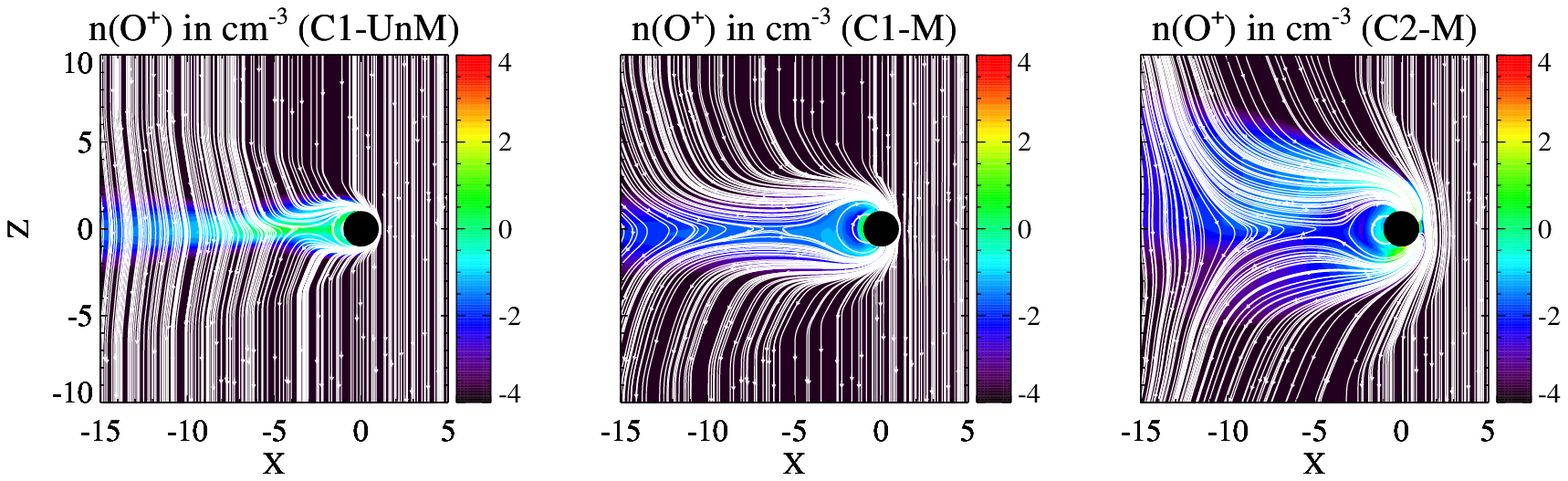}
\includegraphics[scale=1.0]{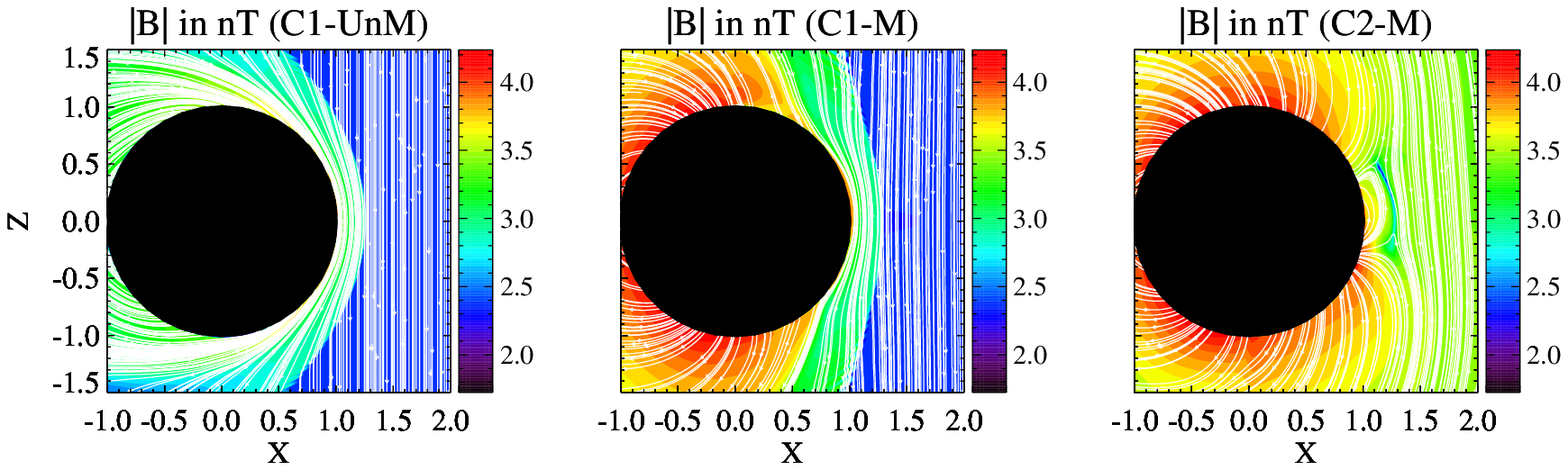}
\caption{The logarithmic scale contour plots of the O$^+$ ion density (first row) and magnetic field strength (second row) with magnetic field lines (in white) in the meridional plane for the unmagnetized case 1 (C1-UnM), magnetized case 1 (C1-M) and magnetized case 2 (C2-M).}\label{BandU}
\end{figure*}

\begin{figure}[!ht]
\centering
\includegraphics[scale=0.5]{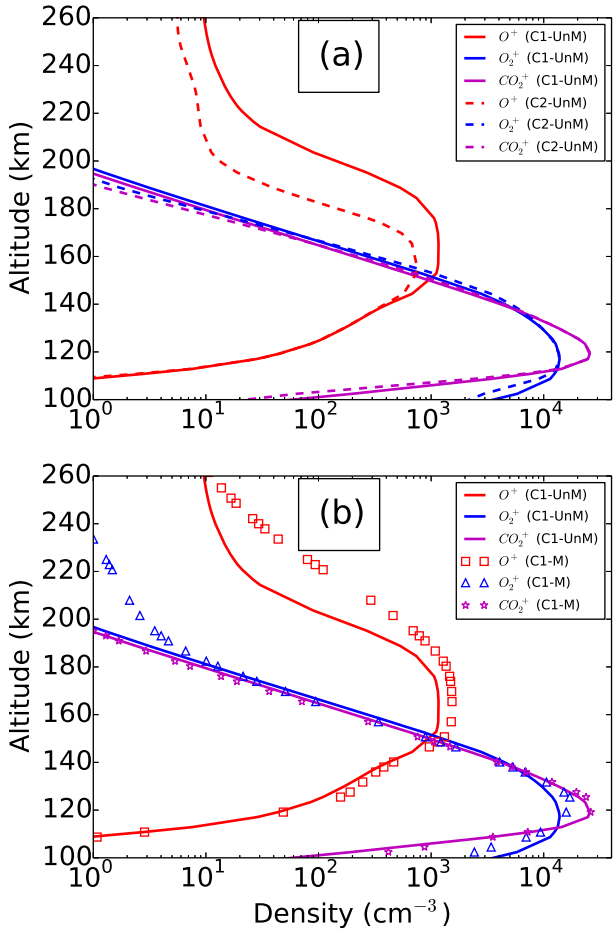}
\caption{The ionospheric profiles along the substellar line for the different cases outlined in Tables \ref{SW} and \ref{table3}.}
\label{iono}
\end{figure}

The most important results of our simulations, i.e., the escape rates (computed by using a sphere with the radius equal to 10 R$_P$) are summarized in Table \ref{table3}. The contour plots of the O$^+$ ion density, the magnetic field strength $\left(|\mathbf{B}|\right)$ and the magnetic field lines for unmagnetized case 1 (C1-UnM), magnetized case 1 (C1-M) and magnetized case 2 (C2-M) are presented in Fig. \ref{BandU}. As seen from Fig. \ref{BandU}, the case with a dipole field is characterized by a relatively large barrier to the stellar wind (from $\mid${\bf B}$\mid$ color contours). In all three instances, the plasma boundaries are greatly compressed, primarily by the stellar wind dynamic pressure, P$_{dyn}$, especially for the magnetized case. The second row in Fig. \ref{BandU} illustrates the plasma boundaries clearly by zooming in. 

The main conclusions are as follows: 1) Ions escape along the open magnetic field lines in all three cases (top row of Fig. \ref{BandU}), from the forces (mostly pressure gradient and $\mathbf{J}$ $\times$ $\mathbf{B}$ terms) in the ion momentum equation. 2) Although C1-M has a global dipole magnetic field, the dipole field strength is not strong enough to fully protect the exoplanet from the stellar wind interaction due to the enormous stellar wind dynamic pressure, and the southward stellar wind B$_z$ that enables dayside magnetic reconnection. 3) Closed dipole field lines are observed for C2-M, and not C1-M, because the value of P$_{dyn}$ for case 2 ($\sim$ 5 $\times$ 10$^4$ nP) is five times smaller than that of case 1 ($\sim$ 2.5 $\times$ 10$^5$ nP). 4) In both C1-M and C2-M, the polar cap regions (i.e., the regions poleward of the auroral oval, \citealt{schunk2009}, pp. 26) are greatly extended to lower latitudes compared to that of the Earth. Thus, the escape rates in both cases are expected to be larger than the typical values observed in our Solar system.

Interestingly, the ionospheric profiles, especially the heavy ions O$_2^+$ and CO$_2^+$, are not significantly affected by the stellar wind conditions $\lesssim$ 200 km (i.e., the photochemical region) regardless of the stellar wind total pressure - see Fig. \ref{iono}. The slight increase in O$^+$ density (in both panels of Fig. \ref{iono}) results from the charge exchange (see Table \ref{table1}) with the enhanced penetrating stellar wind protons (not shown in the plot).

From Table \ref{table3}, we see that the total escape rates for the unmagnetized cases are $\sim 10^{27}$/sec. As expected from Fig. \ref{BandU}, the ion escape rates for C1-UnM and C1-M are similar, although the dipole fields are responsible for reducing the total ion losses to some extent. However, the total ion escape rate for the magnetized case C2-M (compared to C2-UnM) is decreased by about one order of magnitude (mainly caused by decrease in O$^+$) and is on the order of $\sim 10^{26}$/sec. This is consistent with the notion that a global dipole magnetic field protects Venus-like exoplanets from the stellar wind to some extent, although the final rates are still \emph{higher} than those observed in our Solar system. But, this conclusion is not necessarily valid for exoplanets with different sizes, neutral atmosphere profiles and stellar wind parameters. 

In Fig. \ref{iono}, CO$_2^+$ is one of the major ion species in the PCb's ionosphere as a direct consequence of the photoionization of CO$_2$. However, CO$_2^+$ quickly reacts with neutral O to produce mostly O$_2^+$ which becomes the other major ion at a similar altitude as CO$_2^+$. Due to the large scale height of neutral O (because of the lighter mass), the O$^+$ ionospheric peak is located at a relatively high altitude, and is thus picked up first by the stellar wind. Due to momentum conservation, if the stellar wind picks up a large number of light ions, it cannot pick up too many heavy ions. In addition, the electromagnetic shielding caused by the high-altitude ionized O$^+$ ions (e.g, as a conductor layer) also helps in preventing the stellar wind and IMF from penetrating deeply into the planetary ionosphere and picking up the heavy ions. This helps explain why there are \emph{higher} O$^+$ but \emph{lower} heavy ion losses in C1-UnM$_{93}$, as compared to C1-UnM. C1-UnM$_{93}$ is the unmagnetized case with the Venusian surface pressure of $93$ bar instead of the Earth value of $1$ bar with the scale height chosen in accordance with (\ref{ScHPCB}). Overall, the total ion escape rate does not change too much, indicating that the surface pressure does not significantly affect the ion escape rates under extreme stellar wind conditions. 

In contrast to our preceding discussion, it is noteworthy that the terrestrial planets in our Solar system exhibit loss rates that are typically $\sim$10$^{25}$/sec \citep{Lammer13,brain16}.

\section{Discussion and Conclusion}
%We begin by observing that the escape rates for different ions vary considerably, as seen from Table \ref{table3}. Hence, the atmospheric composition and overall mass losses of PCb are expected to change over time.
It was argued in \citet{SEHTM01} that approximately $2\%$ of the atmospheric oxygen had been lost over 3 billion years on Earth. As the escape losses for Proxima Centauri b (PCb) in the \emph{unmagnetized} case are about two orders of magnitude higher,\footnote{Since our choices of PCb's radius and surface atmospheric pressure are close to that of the Earth, the total amount of atmosphere will also be similar to Earth.} all of the atmosphere could be depleted much faster - possibly in a span of $\calo\left(10^8\right)$ years. In turn, this has very important ramifications for surface-based life-as-we-know-it, given the importance of elements such as oxygen \citep{RS06,DecVa10}. 

If gases such as oxygen are depleted on these short timescales, sufficient time may not exist for complex life to evolve. Our simulations indicate that the escape rates for PCb in the \emph{magnetized} case are higher than that of the Earth, implying that some of the above conclusions for the unmagnetized case are also valid here. But, it is equally important to recognize that the magnetized case is quite sensitive to the values of the stellar wind parameters (see Table \ref{table3}). The atmosphere depletion could occur over $\calo\left(10^8\right)$ and $\calo\left(10^9\right)$ years for C1-M and C2-M, respectively.

Thus, our results appear to indicate that PCb, and other similar M-dwarf exoplanets, are not generally capable of supporting an atmosphere over long (Gyr) timescales in both the unmagnetized or magnetized cases when the total stellar wind pressure is \emph{high}. In contrast, for lower values of the pressure, the magnetized case can potentially sustain an atmosphere over Gyr timescales. It has been widely established that the HZ of M-dwarfs evolves over time \citep{SBJ16}. If the stellar wind pressure was higher at earlier epochs, this would have led to increased escape rates. Hence, even with a dipole field, the atmosphere may be eroded on sub-Gyr timescales.

We have also demonstrated that the ionospheric profiles (especially the heavy ions O$_2^+$ and CO$_2^+$) are mostly unaffected by the stellar wind conditions at $\lesssim$ 200 km. This is important because the stellar wind pressure is highly variable, as noted in \citet{GDC16}. Hence, the fact that the stellar wind variability does not greatly influence the lower regions lends some credence to the hypothesis that the surface biology may not be significantly affected \citep{GTSGA}.

There are some important caveats, with regards to the above conclusions, that must be reiterated at this stage. Most of the parameters for PCb and Proxima Centauri are currently unknown. The composition and thickness of the atmosphere could be different, enabling the retention of the atmosphere over longer periods of time. The existence of processes, such as outgassing which occurs on Titan \citep{TLS06}, may counteract the atmospheric losses by serving as \emph{sources}. Thus, a complete understanding of this problem necessitates a knowledge of both source and loss mechanisms. Our model does not include kinetic processes, which contribute to the ion loss mechanisms \citep{strangeway05}.

Finally, as observed in Sec. \ref{SecIntro}, there are $\sim 10^{10}$ exoplanets in the HZ around M-dwarfs. Thus, even if our parameter choices are invalidated by future observations of PCb, these values would, in all likelihood, be valid for other M-dwarf HZ exoplanets whose parameters are similar to the ones considered herein. 
 
To summarize, this Letter sheds light on atmospheric losses from M-dwarf exoplanets in the HZ such as PCb, and clearly delineates the role of the planetary magnetic field and the stellar wind parameters. We have shown that Venus-like exoplanets (with PCb-like parameters) are characterized by high escape rates in the unmagnetized limit, but these values are reduced to an extent when a dipole magnetic field exists. Hence, PCb may undergo significant atmospheric erosion over Gyr timescales in both the magnetized and unmagnetized cases, but this statement must be taken with due caution since there are many inherent uncertainties. Future missions such as JWST will be essential in placing constraints on stellar winds and exoplanet atmospheres, thereby paving the way for more accurate estimations of atmospheric losses. 

\acknowledgments
The authors thank Adam Burrows, Zhenguang Huang, Abraham Loeb, Janet Luhmann and Edwin Turner for the helpful and encouraging comments. This research was supported by the NASA Living With a Star Jack Eddy Postdoctoral Fellowship Program, administered by the University Corporation for Atmospheric Research. Resources supporting this work were provided by the NASA High-End Computing (HEC) Program through the NASA Advanced Supercomputing (NAS) Division at Ames Research Center. The Space Weather Modeling Framework that contains the BATS-R-US code used in this study is publicly available from http://csem.engin.umich.edu/tools/swmf. For distribution of the model results used in this study, please contact the corresponding author. 

\bibliographystyle{aasjournal}
%\bibliography{PCb}

\end{document}